\documentclass[11pt,english]{article}   % MikTeX
\usepackage[russian]{babel}             % MikTeX
\usepackage{amssymb}  % MikTeX
\usepackage[dvips]{graphicx}
\usepackage{amsmath,amssymb}
\usepackage{algorithm}
\usepackage{algorithmic}

\newtheorem{theorem}{Theorem}
\newtheorem{lemma}{Lemma}

\date{}

{\par\addvspace{1mm}{\it Proof\hspace{1.0ex}{#1}.} }%
{\quad$\blacktriangle$\par\addvspace{1mm}}

    \newif\ifNoRemark
    \def\addtheorem#1#2#3#4{ % \usepackage{ifthen} needed
    \ifthenelse{\expandafter\isundefined\csname the#2\endcsname}{\newcounter{#2}}{}
    \newenvironment{#1}[1][\global\NoRemarktrue]% No Remark by default
     {\par\addvspace{2mm}\noindent % ????? ???????? ??? ??????? ??????
       \refstepcounter{#2}{\bf #3~\csname the#2\endcsname
      \vphantom{##1}\ifNoRemark.\ \else\ (##1).\fi}\begingroup #4}%
     {\endgroup\par\addvspace{1mm}\global\NoRemarkfalse}
    \expandafter\newcommand\csname b#1\endcsname{\begin{#1}}
    \expandafter\newcommand\csname e#1\endcsname{\end{#1}}
    }

\begin{document}

\title{A Lower Bound on the Number of Boolean Functions with Median
Correlation Immunity }

\author{
 Vladimir N. Potapov\\ Sobolev Institute of Mathematics,
 Novosibirsk, Russia \\ vpotapov@math.nsc.ru\\ }

\maketitle

\begin{abstract}
 The number of $n$-ary balanced correlation immune
(resilient) Boolean functions of order $\frac{n}{2}$ is not less
than $n^{2^{(n/2)-2}(1+o(1))}$ as $n\rightarrow\infty$.

\end{abstract}

\section{Introduction}

A set $Q_q^n=\{0,1,\dots,q-1\}^n$ with Hamming metric is called an
$n$-dimensional hypercube.  A hypercube is called Boolean if $q=2$.
A subset of $Q^n_q$ consisting of $n$-tuples with fixed  values in
fixed $(n-m)$ coordinates  is called $m$-dimensional face
($m$-face).

%\begin{Definition}
A function $f:Q_q^n\rightarrow \{0,1\}$ is called correlation immune
of order $r$ if it takes the value $1$ the same number of times for
each $(n-r)$-face of the hypercube. A correlation immune function is
called balanced (a resilient function) if it takes  values $0$ and
$1$ the same number of times.
%\end{Definition}

Applications of correlation immune functions in cryptography and
connections between these functions and orthogonal arrays   are
discussed in  \cite{T2}. Further we will investigate resilient
functions. An asymptotic number (as $n\rightarrow\infty$)  of
resilient Boolean functions of order $r=const$ was found in
\cite{Den} and \cite{B}. Methods from \cite{Den} and \cite{B} are
developed in \cite{Can} and \cite{Pan} for the calculation of the
asymptotic number of resilient Boolean functions of order $r=O(n/\ln
n)$. The resilient Boolean functions of order $n-c$, where
$c=const$, are listed in \cite{T1}. The number of resilient Boolean
functions of order   $\alpha n$ for $0<\alpha<1$ remained completely
unknown as $n\rightarrow\infty$. The asymptotic of double logarithm
of the number of such functions is unknown.  In \cite{CK} it is
obtained some upper bound $2^{2^{n-\varepsilon(\alpha)}}$ of the
number of resilient Boolean functions of order  $\alpha n$. This
bound is a bit better then a trivial upper bound  based on an
estimation of algebraic degree of a correlation immune function with
some order. Now we consider the case $\alpha=1/2$
 and obtain a lower bound $n^{2^{(n/2)-1}(1+o(1))}$ for the number of resilient
 functions of order  $\frac{n}{2}-1$. This bound is a bit better then a well-known lower bound $2^{2^{n/2}}$.

\section{Main results}

The lower bound $2^{2^{n/2}}$ follows from a simple construction.
Suppose that $n=2m$. Consider an arbitrary Boolean function
 $f:Q_2^m\rightarrow Q_2$.
Define a function  $F:Q_2^{2m}\rightarrow Q_2$ by the equation
 $F(x,y)=f(x)\oplus |y|$, where $|y|$ is the parity of the Hamming
 weight of  $y$. It is clear that $F$ takes values
 $0$ and $1$ the same number of times in each face with unfixed coordinate $y_i$, $i=1,\dots,m$.
 Consequently,   $F$ is a resilient function of order
$m-1$. The number of such functions is equal to $2^{2^m}$.

In this paper we improve this  bound. At  first we consider
correlation immune functions in $Q^n_4$.  In  \cite{PK} it was found
a sharp asymptotic bound   $3^{n+1}2^{2^n+1}(1+o(1))$  of the number
of
 correlation immune functions $f:Q_4^{n}\rightarrow Q_2$
of order $n-1$ with the frequency  of ones $\frac14$. But we need
the number of
 correlation immune functions $f:Q_4^{n}\rightarrow Q_2$
of order $n-1$ with the frequency  of ones $\frac12$.

\begin{lemma} The number of splittings of $Q^n_2$  into pairwise nonintersecting faces is equal to
${n}^{2^{n-1}(1+o(1))}$ as $n\rightarrow\infty$.
\end{lemma}
Proof. A lower bound follows from the estimation of the numbers of
perfect matchings in Boolean hypercube (see \cite{PP}). Let us prove
an upper bound. Consider a splitting of $Q^n_2$ and arbitrary  face
$L$ from this splitting. If for each vertex in $L$ with even weight
  we have a direction to an arbitrary neighbor  vertex in $L$ with minimum possible weight
   then we can recover this face. Consequently,
 $2^{n-1}$ numbers  from the set $\{1,\dots,n\}$ are sufficient in
order to uniquely indicate the splitting.$\blacktriangle$

Suppose that a splitting of $Q^n_2$ contains $0$-dimensional faces.
It is possible to use number $0$ for an indication that $L$ is a
$0$-dimensional face. In this case the asymptotical bound is the
same because $(n+1)^{2^{n-1}}={n}^{2^{n-1}(1+o(1))}$ as
$n\rightarrow\infty$.

Let us consider $ Q_4^{n}$ as the Cartesian product $ Q_2^{n}\times
Q_2^{n}$. A splitting of $ Q_2^{n}$ into pairwise nonintersecting
faces induces a splitting of $ Q_4^{n}$ into blocks. One part of
coordinates takes two values in a block and another part of
coordinates takes four values in a block. For example, consider the
block $B=\{0,1\}^k\times\{0,1,2,3\}^{n-k}$. Define a function
$f:Q_4^{n}\rightarrow Q_2$ on $B$ by the following equation
$$f|_B(x)=\chi_{1}(x_1)\oplus\dots\oplus\chi_{1}(x_k)\oplus\chi_{2,3}(x_{k+1})
\oplus\dots\oplus\chi_{2,3}(x_{n}),$$ where $\chi_{2,3}$ and
$\chi_{1}$ are indicators of the sets $\{2,3\}$ and $\{1\}$. It is
clear that $f$ is a resilient function of order $n-1$.
% as the parity check function on the
%coordinates from the first part and as another balanced function
%$\delta:Q_4\rightarrow Q_2$ on any coordinate from the second part.
It is not different to verify the following statements.

\begin{lemma} Different splittings of $Q^n_2$ correspond to different resilient functions
$f:Q_4^{n}\rightarrow Q_2$ of order $n-1$.
\end{lemma}

\begin{theorem} There exist at least $n^{2^{(n/2)-1}(1+o(1))}$
different resilient Boolean functions of order $\frac{n}{2}-1$.
\end{theorem}
Proof. Define an arbitrary bijection  $\varphi:Q_2^2\rightarrow
Q_4$. Suppose $f:Q_4^{n}\rightarrow Q_2$ is a resilient function of
order $n-1$. Define function $F:Q_2^{2n}\rightarrow Q_2$ by equation
$F(x,y)=f(\varphi(x_1,y_1),\dots,\varphi(x_n,y_n))$. Let us prove
that $F$ is a resilient Boolean function of order $n-1$. Consider an
arbitrary $(n+1)$-dimensional face $\Gamma$. There exists $i\in
\{1,\dots,n\}$ such that the pair of coordinates $(x_i,y_i)$ is not
fixed in $\Gamma$. Since $f$ takes each of the values $0$ and $1$
two times in any $1$-dimensional face of $Q^n_4$, $F$ takes  each of
the values $0$ and $1$ the same  number of times in $\Gamma$. It is
clear that different resilient functions $f_1$ and $f_2$ correspond
to different resilient functions $F_1$ and $F_2$. So the Theorem 1
follows from Lemmas 1 and 2.$\blacktriangle$

By the simple construction described in the beginning of this
section we can increase together number of variables and correlation
immunity of function.

The work was supported by the program of fundamental scientific
researches of the SB RAS  I.5.1, project No. 0314-2019-0017

\end{document}